\documentclass{SciPost}

\hypersetup{
    colorlinks,
    linkcolor={red!50!black},
    citecolor={blue!50!black},
    urlcolor={blue!80!black}
}

\usepackage[bitstream-charter]{mathdesign}
\DeclareSymbolFont{usualmathcal}{OMS}{cmsy}{m}{n}
\DeclareSymbolFontAlphabet{\mathcal}{usualmathcal}

\begin{document}

\pagestyle{plain}

\begin{center}{\Large \textbf{\color{scipostdeepblue}{
Topological phases of coupled Su-Schrieffer-Heeger wires\\
}}}\end{center}

\begin{center}\textbf{
Anas Abdelwahab\textsuperscript{1$\star$}
}\end{center}

\begin{center}
{\bf 1} Leibniz Universit\"{a}t Hannover, Institute of Theoretical Physics,  Appelstr.~2, 30167 Hannover, Germany
\\[\baselineskip]
$\star$ \href{mailto:anas.abdelwahab@itp.uni-hannover.de}{\small anas.abdelwahab@itp.uni-hannover.de}
\end{center}

\section*{\color{scipostdeepblue}{Abstract}}
\textbf{\boldmath{%
The global half-filled phase diagrams of an arbitrary number $N_{\text{w}}$ of diagonally and perpendicularly coupled Su-Schrieffer-Heeger wires are identified. The diagonally coupled wires exhibit rich topological phase diagrams with insulating phases characterized by winding numbers $0\leq w \leq N_{\text{w}}$ and topological critical lines. They also exhibit completely flat bands at specific values of the model parameters, indicated by lines in the phase diagrams. An even number of perpendicularly coupled wires exhibits either gapless or topologically trivial phases. An odd number of perpendicularly coupled wires additionally exhibits nontrivial topological phases with winding number $w=1$. For
$\delta<0$, their gapless regions contain a topologically nontrivial effective SSH sector supporting localized single-particle edge modes. They also reveal confined coherent correlations in the odd-indexed wires at strong inter-wire coupling away from the gapless regions.
}}

\vspace{\baselineskip}



\vspace{10pt}
\noindent\rule{\textwidth}{1pt}
\tableofcontents
\noindent\rule{\textwidth}{1pt}
\vspace{10pt}


\section{Introduction:}\label{sec:Introduction}
The Su-Schrieffer-Heeger (SSH) model~\cite{su79}, disregarding the harmonic
vibrations, is widely considered one of the simplest models of topological
insulators~\cite{shen12,asb16}. Several variations of it have been studied,
including ladder structures~\cite{zha17,lin17,pad18}, two-dimensional
lattices~\cite{liu17}, chains or wires with long-range hoppings~\cite{maf18}, and
out-of-equilibrium driven SSH models~\cite{maf18,gon19}. Experimentally, the SSH
model has been realized in conjugated polymers~\cite{su79}, trapped neutral atoms
in optical lattices~\cite{ata13,mei16,xie19}, semiconductor quantum
dots~\cite{kic22}, and atom manipulation and design by scanning tunneling
microscopy~\cite{dro17,kha19,hud20,pha22}. Previous investigations have addressed
the analytical solutions and topological properties of coupled SSH
chains~\cite{oba19,kyl20,jan22}, including systems with a few coupled
wires~\cite{zha17,lin17,pad18}, weakly coupled chains~\cite{mat23}, and $N$-
stacked SSH chains studied within a two-dimensional framework~\cite{agr23}.
Nevertheless, to our knowledge, a unified characterization of the global half-
filled phase diagrams for arbitrary finite numbers of diagonally and
perpendicularly coupled SSH wires, together with their critical, gapless, and
correlation properties, has not been presented.

The main advance established by the current work is the exact determination and
physical characterization of the topological, critical, and gapless phases of
coupled SSH wires as quasi-one-dimensional systems. The number of wires is finite,
and the transverse direction is not assigned a Bloch momentum. We provide a
complete identification of the global half-filled phase diagrams for an arbitrary
number of diagonally and perpendicularly coupled SSH wires; see
Fig.~\ref{fig:TopologiaclPhases}. Instead of implementing the details of
transverse-mode decomposition~\cite{del11,zha17,lin17,pad18}, we employ exact
solutions of modified Toeplitz-plus-Hankel matrices~\cite{den21,deng24arxiv} and
construct sets of independent effective two-leg ladders that faithfully represent
the original system. This allows us to determine the phase diagrams exactly. The
modified Toeplitz-plus-Hankel matrices provide compact solutions for a broader
class of systems whose Hamiltonians, or functions thereof, share the same matrix
structure. For the current investigation, they produce the transverse-resolved
solutions for the coupled SSH wires. The gapped phases are characterized by the
BDI winding number of the full finite-width Hamiltonian. The resulting phase
diagrams reveal the spectral structure of the coupled wires, including the
emergence of flat bands, critical and gapless phases, and the associated edge-
state and correlation properties of the half-filled ground state.

This work emphasizes the SSH model not only as a typical model of topological
insulators, but also as a model of topological critical and gapless systems.
Therefore, we employ the topological characterization introduced by Verresen
\textit{et al.}~\cite{ver18} to investigate the critical and gapless phases. Our
results reveal important qualifications concerning the application of their edge-
mode characterization to systems with additional transverse symmetries. In
particular, diagonally coupled wires can exhibit critical points without localized
edge modes between insulating phases with different positive winding numbers,
while odd numbers of perpendicularly coupled wires can support localized edge
modes in gapless phases even when the full-system topological index vanishes or
becomes negative.

Finally, for odd numbers of perpendicularly coupled wires, we find confined edge
states with probabilities equally distributed among odd-indexed wires, together
with coherent long-distance single-particle correlations along these wires at
sufficiently strong inter-wire coupling. The edge-state probability and the
nonlocal correlations vanish in the wires with even indices. The corresponding
edge state resembles a W-like state~\cite{dur00}, in the sense of coherent single-
particle superposition across spatial modes, rather than genuine multipartite
entanglement between particles.

We emphasize that the distinction between mode-resolved SSH topology and the
global half-filled coupled-wire phase structure is essential, especially for
interpreting the gapless regimes and the W-like correlation patterns. The approach
used in this work is based on the exact decomposition of the full coupled SSH-wire
systems into sets of effective two-wire subsystems, supplemented by a single
effective wire when the total number of wires is odd. The global half-filled
topological phases can then be constructed straightforwardly. A similar approach
has been used for coupled Hubbard wires~\cite{led00}. An important motivation for
the present investigation is the established correspondence between the
topological phases of the single SSH chain and the symmetry-protected topological
(SPT) phases of the SSH-Hubbard chain at half filling and the dimerized spin-$1/2$
Heisenberg chain~\cite{hid92,man12,mik22}. The current work provides a
noninteracting reference for future investigations of how the topological phases
evolve under interactions in coupled spinless fermion, spinful fermion, and spin
systems, and whether analogous correspondences exist for coupled wires.
Furthermore, it is interesting to investigate the fate of the W-like edge states
and confined coherent correlations under interactions. The competition between
dimerization, inter-wire coupling, and interactions may lead to richer phase
diagrams and different SPT phases.

\section{The models}\label{sec:Models}
\begin{figure*}[t]
\centering
\includegraphics[width=0.4\textwidth]{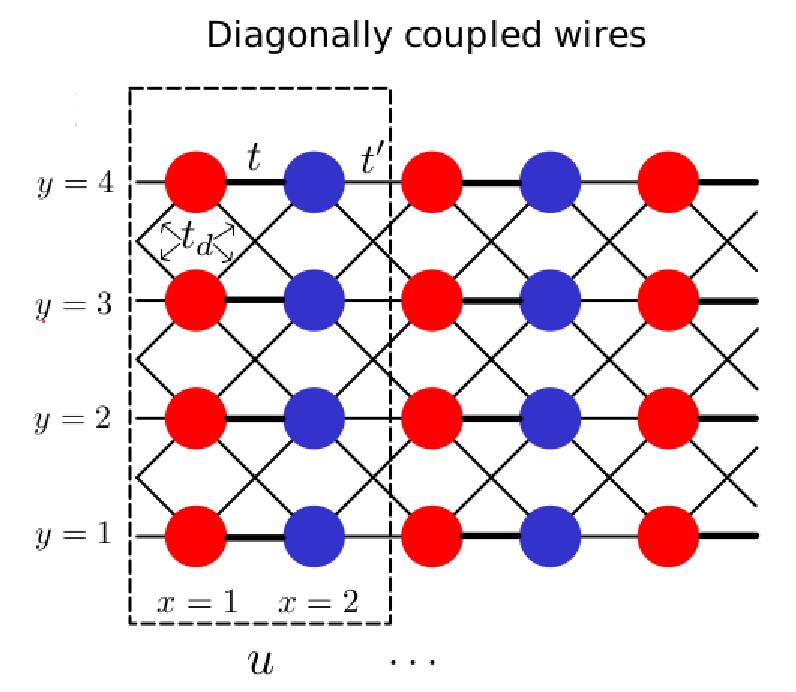}
\includegraphics[width=0.4\textwidth]{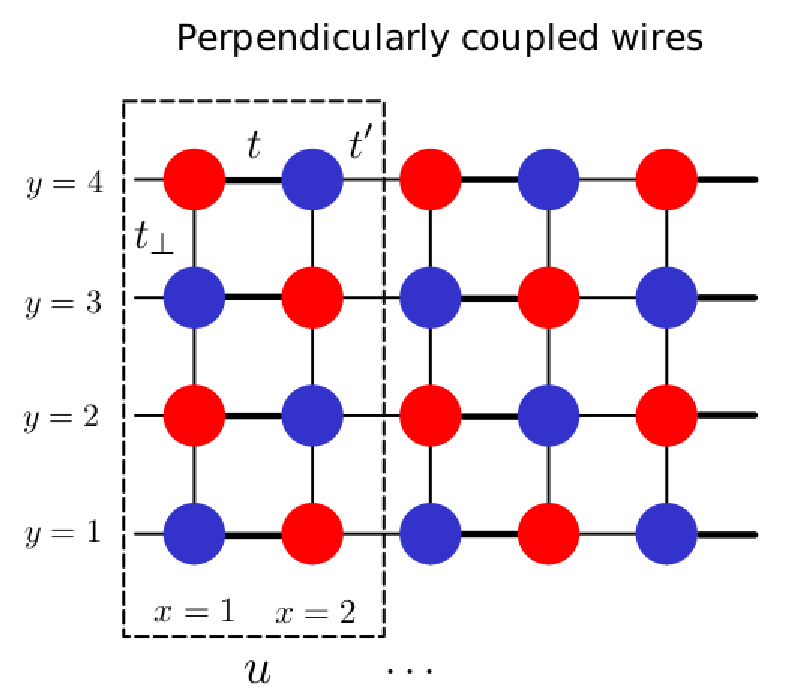}
\caption{Lattice structures of diagonally and perpendicularly coupled wires.}
\label{fig:LadderStrucures}
\end{figure*}
The diagonally and perpendicularly coupled SSH wires shown in Fig.~\ref{fig:LadderStrucures} are described by the Hamiltonian
\begin{equation}\label{eq:FullHamiltonian}
 H = \sum_{y=1,...,N_{\text{w}}} H_{y} + \sum_{y=1,...,N_{\text{w}-1}} H_{y,y+1} .
\end{equation}
Here, $H_{y}$ represents the $y$th single SSH wire, given by
\begin{equation}\label{eq:HamiltonianSSH}
H_{y} =  \sum_{u} t \left( c^{\dag}_{u,1,y}
c^{\phantom{\dag}}_{u,2,y} + \text{H.c.} \right)
+ \sum_{u} t^{\prime} \left( c^{\dag}_{u,2,y}
c^{\phantom{{\dag}}}_{u+1,1,y} + \text{H.c.} \right) .
\end{equation}
The diagonal coupling is given by
\begin{eqnarray}\label{eq:diagonal-hopping}
 H_{y,y+1} &=&  \sum_{u} t_d \left( c^{\dag}_{u,1,y} c^{\phantom{{\dag}}}
 _{u,2,y+1} + \text{H.c.} \right)
+  \sum_{u} t_d \left( c^{\dag}_{u,2,y} c^{\phantom{{\dag}}}
_{u,1,y+1} + \text{H.c.} \right)
\nonumber \\
&+&  \sum_{u} t_d \left( c^{\dag}_{u,2,y} c^{\phantom{{\dag}}}
_{u+1,1,y+1} + \text{H.c.} \right)
+ \sum_{u} t_d \left( c^{\dag}_{u,2,y+1} c^{\phantom{{\dag}}}
_{u+1,1,y} + \text{H.c.} \right) .
\end{eqnarray}
The perpendicular coupling is given by
\begin{equation}\label{eq:perpendicular-hopping}
 H_{y,y+1} =  \sum_{u} t_{\perp} \left( c^{\dag}_{u,1,y}
 c^{\phantom{{\dag}}}_{u,1,y+1} + \text{H.c.} \right)
+  \sum_{u} t_{\perp} \left( c^{\dag}_{u,2,y}
c^{\phantom{{\dag}}}_{u,2,y+1} + \text{H.c.} \right)
\end{equation}
Here, $c^{\dag}_{u,x,y}$ ($c^{\phantom{{\dag}}}_{u,x,y}$) denotes 
the creation (annihilation) operator of a spinless fermion in unit 
cell $u$. The coordinates $(x,y)$ are restricted to the unit cell 
$u$. The total number of unit cells is $N_u$, and the total number 
of wires is $N_{\text{w}}$. While $x=1,2$ labels the sites within a 
unit cell, we use $r=2u-x\%2=1,2,3,\ldots,L_x$ to denote the 
bare rung index, where $L_x$ is the total number of rungs. Unless 
explicitly stated, we always consider even $L_x$. The intra-wire 
hopping parameters are given by $t=1+\delta$ and $t^{\prime}=1-\delta$, with dimerization $-1 \leq \delta \leq 1$. The inter-wire 
hoppings are given by $t_d>0$ for diagonally coupled wires and 
$t_{\perp}>0$ for perpendicularly coupled wires. Periodic boundary 
conditions (PBC) along the wire direction correspond to finite $N_u$ 
such that $c^{\dag}_{1,x,y}=c^{\dag}_{N_u+1,x,y}$. Open boundary 
conditions (OBC) along the wire direction correspond to finite $N_u$ 
such that the wires terminate at $u=1$ and $(1,y)$, as well as at 
$u=N_u$ and $(2,y)$.

The coupled SSH wires belong to the BDI 
class~\cite{shen12,asb16,sch08ClassTI,sch09ClassTI,kitaevTI,ryu10}. 
Therefore, due to chiral symmetry, there exists a Hermitian unitary 
operator $\Gamma$ that anticommutes with $H$, i.e., $\Gamma H \Gamma = -H$. As a consequence, the system can be divided into two 
sublattices such that the Hamiltonian connects only sites belonging 
to different sublattices. The two sublattices are distinguished by 
the blue and red colors in Fig.~\ref{fig:LadderStrucures}. 
Accordingly, one can define canonical bases such that the first 
(second) half of their elements corresponds to the first (second) 
sublattice. Then, $H$ can be written in a completely block-off-diagonal form, while the chiral operator takes the diagonal form
\begin{equation}
\label{eq:GammaDiagonalForm}
 \Gamma =
\begin{bmatrix}
\mathbb{1} & 0\\
0 & -\mathbb{1}
\end{bmatrix} ,
\end{equation}

$H$ can be written as a sum of commuting operators $H(k)$ acting only on single-particle Bloch states with wave number $k$ in the first Brillouin zone. The Bloch states can be written in canonical forms such that $H(k)$ inherits the chiral properties and can be expressed in the off-diagonal form
\begin{equation}
\label{eq:OffDiagonalForm}
 H(k) =
\begin{bmatrix}
0& h^{\dagger}(k)\\
h(k)& 0
\end{bmatrix} ,
\end{equation}
where $H(k)$ is a $2N_{\text{w}}\times 2N_{\text{w}}$ matrix, while $h(k)$ and $h^{\dagger}(k)$ are $N_{\text{w}}\times N_{\text{w}}$ matrices.
The block-off-diagonal matrix $h(k)$ of the coupled SSH wires takes the general tridiagonal form
\begin{equation}\label{eq:OffDiagonalGeneralForm}
h(k) =
 \begin{bmatrix}
    b & c & 0 & 0 & \hdots \\
    c & a & c & 0 &  \hdots \\
    0 & c & b & c &  \hdots \\
    0 & 0 & c & a &  \ddots \\
    \vdots & \vdots     & \vdots & \ddots & \ddots
\end{bmatrix} ,
\end{equation}
where $a$, $b$, and $c$ are complex entries that depend on the model, as discussed below.

The topological insulating phases are characterized by the winding number $w\in\mathbb{Z}$, obtained using~\cite{shen12,asb16,ber18,maf18}
\begin{equation}\label{eq:WindingNumber}
     w = \frac{1}{2i\pi} \sum^{N_{\text{w}}}_{\lambda=1} \int^{\pi}_{-\pi}
     \frac{\partial}{\partial k} \text{log}\left[h_{\lambda}(k)\right] dk ,
\end{equation}
where $h_{\lambda}(k)$ denotes the complex eigenvalues of the off-diagonal block $h(k)$. The winding number $w$ is the winding number of the trajectory generated by the product $\prod^{N_{\text{w}}}_{\lambda=1}h_{\lambda}(k)$ around the origin of the complex plane for $k\in[-\pi,\pi)$. When the gap closes, this trajectory passes through the origin, and the winding number becomes ill defined. The quantity $|w|$ gives the number of exponentially localized edge states at energy $E=0$ for coupled wires with OBC in the thermodynamic limit.

The square of $H(k)$~\cite{maf18} is
\begin{equation}
\label{eq:SquareH}
 H^2(k) =
\begin{bmatrix}
\bar{H}(k)& 0 \\
0& \tilde{H}(k)
\end{bmatrix} ,
\end{equation}
where
$\bar{H}(k)=h^{\dagger}(k)h(k)$ and $\tilde{H}(k)=h(k)h^{\dagger}(k)$
take the form of pentadiagonal banded matrices. We denote the eigenvectors of $\bar{H}(k)$ and $\tilde{H}(k)$ by the sets $\left\lbrace|\bar{\phi}_{\mu}(k)\rangle\right\rbrace$ and $\left\lbrace|\tilde{\phi}_{\mu}(k)\rangle\right\rbrace$, respectively. Then, the energy eigenvectors $|\phi_{\mu}(k)\rangle$ of $H(k)$ are divided into two chiral subsets. For eigenvalues $E_{\mu}(k)>0$, the first subset is given by the direct sum
$|\phi_{\mu}(k)\rangle_{+E}=|\bar{\phi}_{\mu}(k)\rangle \oplus |\tilde{\phi}_{\mu}(k)\rangle$. Thus, using the chiral operator~(\ref{eq:GammaDiagonalForm}), we obtain $|\phi_{\mu}(k)\rangle_{-E}=\Gamma|\phi_{\mu}(k)\rangle_{+E}=|\bar{\phi}_{\mu}(k)\rangle \oplus -|\tilde{\phi}_{\mu}(k)\rangle$ for $E_{\mu}(k)<0$. For zero-energy eigenvalues, we use the relations
\begin{subequations}\label{eq:Coupled_h_hdag}
 \begin{equation}
  h^{\dagger}(k)|\tilde{\phi}_{\mu}(k)\rangle = E(k)|\bar{\phi}_{\mu}(k)\rangle
 \end{equation}
  \begin{equation}
  h(k)|\bar{\phi}_{\mu}(k)\rangle = E(k)|\tilde{\phi}_{\mu}(k)\rangle
 \end{equation}
\end{subequations}
together with the fact that $|\bar{\phi}_{\mu}(k)\rangle$ and $|\tilde{\phi}_{\mu}(k)\rangle$ correspond to different sublattices. Hence, at $E(k)=0$, $|\tilde{\phi}_{\mu}(k)\rangle$ becomes a kernel of $h^{\dagger}(k)$ in Eq.~(\ref{eq:Coupled_h_hdag})a, with $|\bar{\phi}_{\mu}(k)\rangle=0$ in one sublattice. Similarly, $|\tilde{\phi}_{\mu}(k)\rangle=0$ in the other sublattice, while $|\bar{\phi}_{\mu}(k)\rangle$ becomes a kernel of $h(k)$ in Eq.~(\ref{eq:Coupled_h_hdag})b.

$\bar{H}(k)$ and $\tilde{H}(k)$ take the form of so-called modified 
Toeplitz-plus-Hankel matrices, for which exact analytical solutions 
are available~\cite{den21,deng24arxiv}. This allows a complete 
identification of the band structures of $H(k)$. Once the band 
structures are identified, one can construct sets of effective 
independent two diagonally/perpendicularly coupled SSH wires, which 
in turn allow an exact identification of the full topological phase 
diagrams.

\section{Band structures and decomposition into independent effective two-wire subsystems}\label{sec:BandStructures}
\begin{figure*}[t]
\centering
\includegraphics[width=0.4\textwidth]{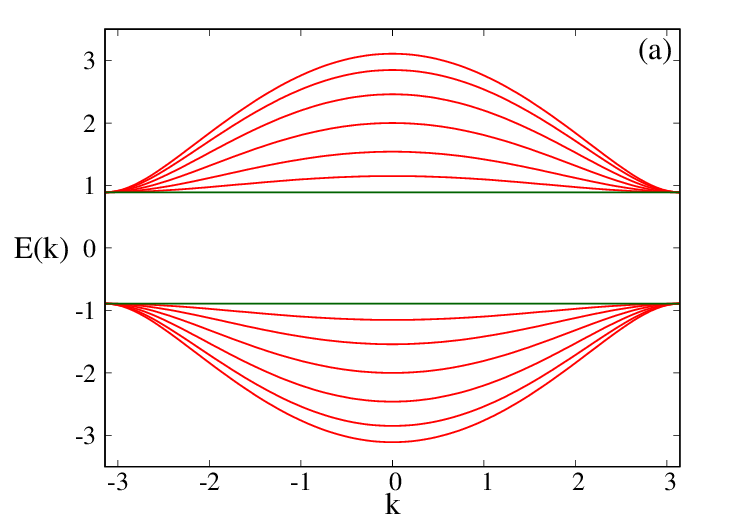}
\includegraphics[width=0.4\textwidth]{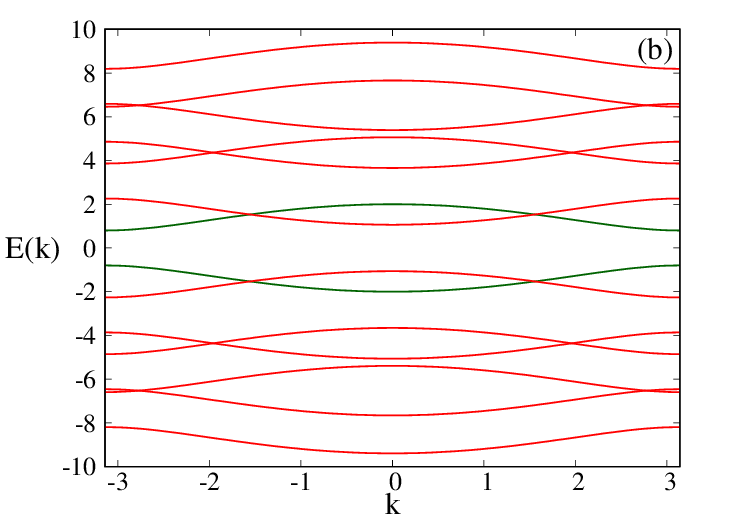}
\caption{Band structures of seven coupled SSH wires for (a) diagonally coupled wires with $t_d=0.3$ and $\delta=\pm(t^{l=1}_{d}-1)$, with clearly visible flat bands, and (b) perpendicularly coupled wires with $\delta=\pm0.4$ and $t_{\perp}=4$.}
\label{fig:BandStructures}
\end{figure*}
Before presenting the analytical results, we briefly comment on previous approaches based on transverse modes. The diagonalization of coupled SSH wires in terms of transverse modes, and the resulting decomposition into effective subsystems, has been discussed previously in various forms~\cite{oba19,kyl20}. In the present work, however, this decomposition is not taken as a starting point, but rather emerges naturally from the exact analytical solutions of the matrices $h(k)$, $h^{\dagger}(k)$, $\bar{H}(k)$, and $\tilde{H}(k)$, using the results for tridiagonal and modified Toeplitz-plus-Hankel matrices~\cite{den21,deng24arxiv}. These solutions provide a compact approach to obtaining the band structures and eigenvectors needed for investigating the physical properties of the coupled wires.

We exploit these exact solutions to determine the band structures and, consequently, the global half-filled topological phase diagrams for an arbitrary number of coupled wires. The decomposition into a set of independent effective subsystems labeled by $l$ provides a convenient and exact framework not only for organizing the resulting phases but also for understanding their topological, edge-state, and correlation properties.

For diagonally coupled SSH wires, $a=b=T(k)=t+t^{\prime}\exp(ik)$ and $c=T_d(k)=t_d+t_d\exp(ik)$. Thus, $h(k)$ is a regular tridiagonal matrix with complex eigenvalues~\cite{den21}
\begin{equation}\label{eq:hEigenValsDiag}
 h_{\mp l}(k) = T(k) \mp 2 T_d(k) \cos\left(\frac{l\pi}{N_{\text{w}}+1}\right) ,
\end{equation}
where $l=1,...,N$ and $N=\frac{N_{\text{w}}}{2}$ $\left(N=\frac{N_{\text{w}}-1}{2}\right)$ for an even (odd) number of wires. For an odd number of wires, the remaining complex eigenvalue is $h_{l_0}(k)=T(k)$, where $l_0=\frac{N_{\text{w}}+1}{2}$. The pentadiagonal matrix $\bar{H}(k)$ can be solved according to~\cite{den21}. It takes the form
\begin{equation}\label{eq:hdaghDiagSSH}
\bar{H}(k) =
 \begin{bmatrix}
    A_d(k)-|T_d(k)|^2 & B_d(k) & |T_d(k)|^2 & 0 & \hdots \\
    B_d(k) & A_d(k) & B_d(k) & |T_d(k)|^2 & \hdots \\
    |T_d(k)|^2 & B_d(k) & A_d(k) & B_d(k) & \ddots \\
    0 & |T_d(k)|^2 & B_d(k) & A_d(k) & \ddots \\
    \vdots & \vdots & \ddots & \ddots & \ddots
\end{bmatrix} .
\end{equation}
Here, $A_d(k)=|T(k)|^2+2|T_d(k)|^2$ and $B_d(k)=T(k)T^{*}_d(k)+T^{*}(k)T_d(k)$. The eigenvalues are
\begin{equation}\label{eq:EkDiag}
 E^2_{\mp l}(k) = t^{2}_{\mp l}+t^{\prime2}_{\mp l}
 +2t^{\phantom{.}}_{\mp l}t^{\prime}_{\mp l}\cos(k) ,
\end{equation}
with hopping parameters $t^{\phantom{.}}_{\mp l}=t\mp t^{l}_d$, $t^{\prime}_{\mp l}=t^{\prime}\mp t^{l}_d$, and
$t^{l}_d=2t_d\cos\left(\frac{l\pi}{N_{\text{w}}+1}\right)$. Note that $\cos\left(\frac{\lambda\pi}{N_{\text{w}}+1}\right)$ with $\lambda=1,...,N_{\text{w}}$ is replaced by $\mp\cos\left(\frac{l\pi}{N_{\text{w}}+1}\right)$ with $l=1,...,N$. The square roots of the eigenvalues in Eq.~(\ref{eq:EkDiag}) give the band structure of $H(k)$. For each $\mp l$, there are two bands resembling those of an effective single SSH wire.

Following~\cite{son24}, the bands corresponding to each $l$ index represent bands of independent effective two diagonally coupled SSH wires, coupled similarly to Eq.~(\ref{eq:diagonal-hopping}), but with $t^{l}_d$ instead of $t_d$. For an odd number of diagonally coupled SSH wires, the remaining eigenvalue $E^2_{l_0}(k)$ gives two bands of a single SSH wire,
\begin{equation}\label{eq:SingleSSHbands}
 E_{l_0}(k) = E(k) = \pm \sqrt{t^{2}+t^{\prime2} +2t^{\phantom{.}}t^{\prime}\cos(k)} .
\end{equation}
We emphasize that this decomposition follows directly from the exact spectrum and does not rely on an approximate projection or weak-coupling argument. Figure~\ref{fig:BandStructures}(a) shows the band structure of seven diagonally coupled wires.

The eigenvectors $|\bar{\phi}_{\mp l}(k)\rangle$ in the canonical basis are given, up to a normalization factor, by~\cite{den21}
\begin{equation}\label{eq:EigenStatesDiag_barphi_mpl}
|\bar{\phi}_{\mp l}(k)\rangle=
 \begin{pmatrix}
\bar{\phi}_{\mp l,1} & \bar{\phi}_{\mp l,2} & \bar{\phi}_{\mp l,3} &  \hdots
\end{pmatrix}^{\mathrm T}
\end{equation}
where $\bar{\phi}_{l,y} = \sin\left( \frac{ly\pi}{N_{\text{w}}+1} \right)$ and $\bar{\phi}_{-l,y} = \left(-1\right)^{y+1}\sin\left( \frac{ly\pi}{N_{\text{w}}+1} \right)$. Similarly, the eigenvectors $|\bar{\phi}_{l_0}(k)\rangle$ are given by
\begin{equation}\label{eq:EigenStatesDiag_barphi_l0}
|\bar{\phi}_{l_0}(k)\rangle=
 \begin{pmatrix}
\bar{\phi}_{l_0,1} & \bar{\phi}_{l_0,2} & \bar{\phi}_{l_0,3} &  \hdots
\end{pmatrix}^{\mathrm T}
\end{equation}
where $\bar{\phi}_{l_0,y} = \sin\left( \frac{y\pi}{2} \right)$. We observe that all even components of $|\bar{\phi}_{l_0}(k)\rangle$ are zero, while the odd components alternate between $1$ and $-1$. This will lead to interesting consequences for the correlation function in perpendicularly coupled wires, as we discuss in Sec.~\ref{sec:ZeroEdgeModes}. To obtain the eigenvectors of $\tilde{H}(k)$, the coupled equations~(\ref{eq:Coupled_h_hdag}a) and~(\ref{eq:Coupled_h_hdag}b) impose important constraints that introduce $k$-dependent phases into $|\tilde{\phi}_{\mp l}(k)\rangle$ and $|\tilde{\phi}_{l_0}(k)\rangle$, despite their absence in Eqs.~(\ref{eq:EigenStatesDiag_barphi_mpl}) and~(\ref{eq:EigenStatesDiag_barphi_l0}).

Since the exact decomposition, for diagonally coupled wires, factorizes $\det h(k)$ into the product of the contributions from the independent $l$- and $l_0$-labeled subsystems, the winding number of the full system, when gapped at zero energy, is the sum of the winding numbers of these subsystems. Notice that each independent sector has to fulfill all the symmetries of class BDI to be classified within this symmetry class independently. This is fulfilled for all $\mp l$ and $l_0$ sectors of the diagonally coupled wires. This gives rise to a rich phase diagram, as discussed below.

For perpendicularly coupled SSH wires, $b=a^{*}=T(k)=t+t^{\prime}\exp(ik)$ and $c=t_{\perp}$. According to~\cite{den21,deng24arxiv}, the eigenvalues are
\begin{equation}\label{eq:hEigenValsPerp}
 h_{\mp l}(k) = \frac{1}{2}\left( F(k) \mp \sqrt{F^2(k) - 4G(k) +
 8t^2_{\perp} \Lambda}  \right),
\end{equation}
where $F(k)=T(k)+T^{*}(k)$, $G(k)=T(k)T^{*}(k)$, and $\Lambda= 1 +
\cos\left(\frac{2l\pi}{N_{\text{w}}+1}\right)$. The product of each $\mp l$ pair is a real number. However, for an odd number of legs, the remaining eigenvalue is given by $h_{l_0}(k)=T(k)$, which carries exclusively the winding number of the full system when the latter is gapped at zero energy.

The pentadiagonal matrix $\bar{H}(k)$ can be solved according to~\cite{deng24arxiv}. It takes the form
\begin{equation}\label{eq:hdaghPerpSSH}
\bar{H}(k) =
 \begin{bmatrix}
    A_{\perp}(k)-t^2_{\perp} & B_{\perp}(k) & t^2_{\perp} & 0 & \hdots \\
    B^{*}_{\perp}(k) & A_{\perp}(k) & B^{*}_{\perp}(k) & t^2_{\perp} &
    \hdots \\
    t^2_{\perp} & B_{\perp}(k) & A_{\perp}(k) & B_{\perp}(k) & \ddots \\
    0 & t^2_{\perp} & B^{*}_{\perp}(k) & A_{\perp}(k) & \ddots \\
    \vdots & \vdots & \ddots & \ddots & \ddots
\end{bmatrix} .
\end{equation}
Here, $A_{\perp}(k)=|T(k)|^2+2t^2_{\perp}$ and $B_{\perp}(k)=2T^{*}(k)t_{\perp}$. The eigenvalues of these matrices are~\cite{deng24arxiv}
\begin{equation}\label{eq:EEkPerp}
 E^2_{\mp l}(k) = A_{\perp}(k) + 2\cos\left(\frac{2l\pi}
 {N_{\text{w}}+1}\right) t^{2}_{\perp} 
 \mp 2\cos\left(\frac{l\pi}{N_{\text{w}}+1}\right)|B_{\perp}(k)| .
\end{equation}
This solution can be straightforwardly reduced to the band structures
\begin{equation}\label{eq:EkPerp}
 E_{\mp l}(k) = E(k) \mp t^{l}_{\perp} ,
\end{equation}
where $t^{l}_{\perp} = 2\cos\left(\frac{\pi l}{N_{\text{w}}+1}\right)t_{\perp}$. Note that for each $\mp l$ there are two bands resembling those of an effective single SSH wire, shifted by an onsite chemical potential given by $\mp t^{l}_{\perp}$, respectively. For an odd number of perpendicularly coupled SSH wires, the remaining two bands again represent an effective single SSH wire with band structure given by Eq.~(\ref{eq:SingleSSHbands}). Figure~\ref{fig:BandStructures}(b) shows the band structure of seven perpendicularly coupled wires.

Following~\cite{son24}, we obtain a set of independent effective two perpendicularly coupled SSH wires, each labeled by $l$. Their perpendicular coupling is similar to that in Eq.~(\ref{eq:perpendicular-hopping}), but with effective perpendicular hopping $t^{l}_{\perp}$ instead of $t_{\perp}$. While the effective perpendicularly coupled wires can acquire only gapless or trivially gapped phases~\cite{son24}, the effective single SSH wire can also exhibit a nontrivial gapped phase. This observation has interesting consequences for the topological classification of the gapless phases, as discussed in Sec.~\ref{sec:Critical_and_Gapless}.

The vector components of the eigenvectors $|\bar{\phi}_{\mp l}(k)\rangle$ for the perpendicularly coupled wires are given in the canonical basis by~\cite{deng24arxiv}
\begin{equation}\label{eq:EigenStatesPerp}
 \bar{\phi}_{+l,y}(k) = \sin\left( \frac{ly\pi}{N_{\text{w}}+1} \right)\begin{cases}
                        1 \text{\hspace{16mm} if $y \in$ odd} \\
                        \frac{B^{*}_{\perp}(k)}{\left|B_{\perp}(k)\right|} \text{\hspace{9mm} if $y \in$ even} ,
                    \end{cases}
\end{equation}
while $\bar{\phi}_{-l,y}(k) = \left(-1\right)^{y+1}\bar{\phi}_{+l,y}(k)$ and $|\bar{\phi}_{l_0}\rangle$ is exactly similar to the corresponding eigenvector obtained for diagonally coupled wires. The eigenvector structure follows from Ref.~\cite{deng24arxiv}, with the factor $B_\perp^*(k)/|B_\perp(k)|$ accounting for the relative phase between odd- and even-indexed components required by the complex off-diagonal elements of $\bar H(k)$ in the canonical basis. The eigenvector expressions involving these relative phases are understood away from the corresponding band degeneracies at $T(k)=0$, where the eigenvectors are not uniquely determined. Moreover, $|\tilde{\phi}_{\mp l}(k)\rangle$ and $|\tilde{\phi}_{l_0}(k)\rangle$ are likewise constrained by the coupled equations~(\ref{eq:Coupled_h_hdag}). Notice that each of the $l$ subsystems fulfills the BDI class symmetries but each of the $\mp l$ subsystems break the particle-hole symmetry. Therefore, the decomposition into two-wire subsystems allows their independent classification within class BDI.

\section{The phase diagrams}\label{sec:PhaseDiagrams}
\begin{figure*}[t]
\centering
\includegraphics[width=0.4\textwidth]{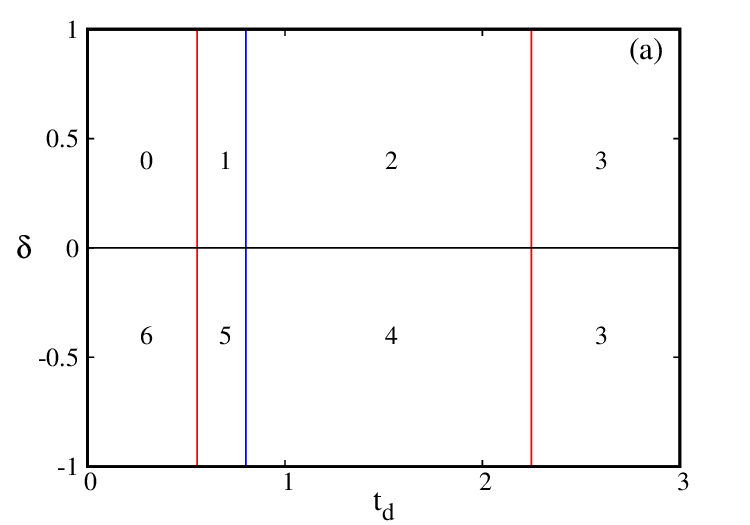}
\includegraphics[width=0.4\textwidth]{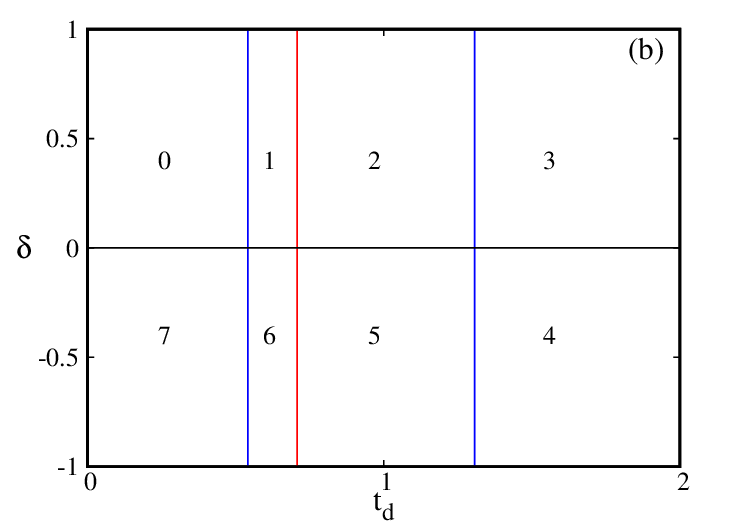}
\includegraphics[width=0.4\textwidth]{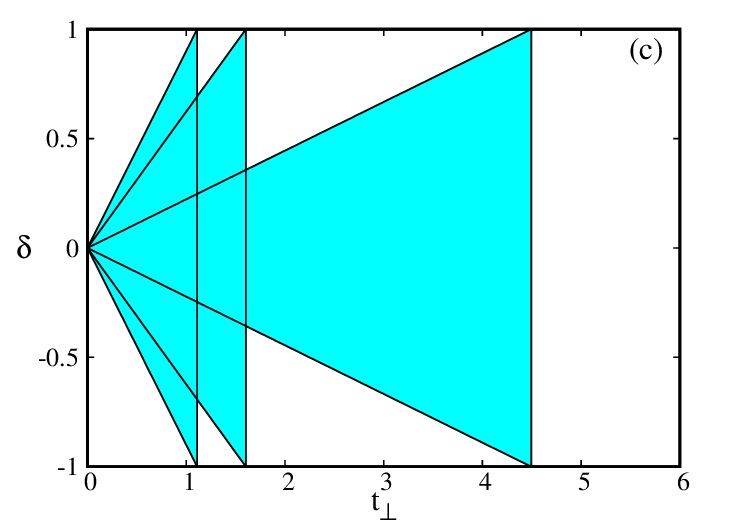}
\includegraphics[width=0.4\textwidth]{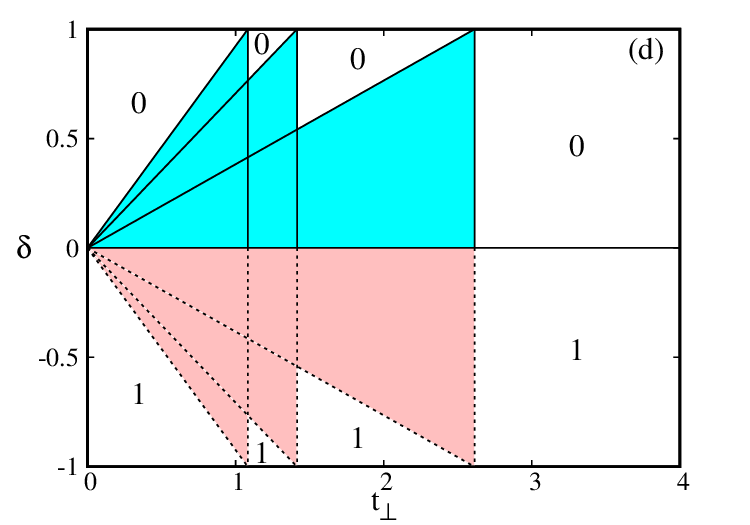}
\caption{Global half-filled phase diagrams of coupled SSH wires (a) and (b) show the phase diagrams of six and seven diagonally coupled wires, respectively. The vertical red (blue) solid lines indicate critical values $\tau^l_d$. (c) and (d) show the phase diagrams of six and seven perpendicularly coupled wires, respectively. The transparent blue regions in (c) indicate gapless phases, while the transparent blue (pink) regions in (d) indicate gapless phases in which the effective single SSH wire is trivial (nontrivial).}
\label{fig:TopologiaclPhases}
\end{figure*}
The construction of independent effective two coupled wires labeled by $l$, together with the effective single SSH wire labeled by $l_0$ for an odd number of wires, allows the determination of the phase diagrams for an arbitrary number of diagonally or perpendicularly coupled SSH wires. This is achieved by identifying the phase diagram of each $l$- and $l_0$-labeled subsystem, following Ref.~\cite{son24}. These individual phase diagrams can then be combined into a single overall phase diagram corresponding to a given number of wires $N_{\text{w}}$. In the following, we discuss these overall phase diagrams.

Each effective two diagonally coupled wire labeled by $l$ has a critical diagonal hopping
$t_d=\tau^l_d=\left[2\cos\left(\frac{l\pi}{N_{\text{w}}+1}\right)\right]^{-1}$, i.e., a hopping amplitude at which the band gap between the $-l$-labeled bands closes. Therefore, for any fixed $\delta\neq0$, the full system undergoes $N$ phase transitions as $t_d$ is increased from zero. By crossing the $\delta=0$ line for any fixed $t_d$ away from the vertical critical lines, all trivial (nontrivial) bands become nontrivial (trivial). For an odd number of diagonally coupled wires, the effective single SSH wire labeled by $l_0$ undergoes a phase transition only at $\delta=0$.

Figures~\ref{fig:TopologiaclPhases}(a) and (b) show the topological phase diagrams of six and seven diagonally coupled SSH wires, respectively. The integers inside the figures give the winding numbers of the insulating phases. The vertical lines are critical lines corresponding to the three possible values of $\tau^l_d$. From the band structure in Eq.~(\ref{eq:EkDiag}), we deduce that the $E_{-l}(k)$ bands become completely flat at
$ \delta = \pm \left( t^{l}_d - 1 \right)$. Note that the $-l$ bands with $ \delta = + \left( t^{l}_d - 1 \right)$ $\left( \delta = - \left( t^{l}_d - 1 \right)\right)$ are nontrivial (trivial). This is an interesting finding, since in possible experimental realizations the system may become susceptible to strong interaction effects at these parameter values, which could lead to exotic quantum states~\cite{bis11,cao18}.

However, the flatness of the bands at $\delta = \pm \left( t^{l}_d - 1 \right)$
is not protected by a symmetry, but rather results from a fine-tuned
cancellation of hopping amplitudes in the effective $l$-labeled
subsystems. Consequently, perturbations that modify this interference
condition generally lift the exact flatness of the bands. Depending on
the nature of the perturbation, this can lead either to dispersive bands
or to a broadening and localization of the corresponding states.
Nevertheless, the topological character of the bands, as determined by
the winding number, remains well defined as long as the symmetries of
the BDI class are preserved and the gap does not close.

The phase diagram of each $l$-labeled effective two perpendicularly coupled wire consists of a triangularly shaped gapless region~\cite{son24}. When more than one effective two-wire subsystem is present, each such region intersects with the gapless regions of other $l$-labeled effective two perpendicularly coupled wires. Each triangle is bounded by the vertical line corresponding to the critical perpendicular hopping
$t_{\perp}=\tau^l_{\perp}=\left[\cos\left(\frac{l\pi}{N_{\text{w}}+1}\right)\right]^{-1}$.
The other two boundaries are the lines
$\delta=\pm\frac{1}{\tau^l_{\perp}}t_{\perp}$. For an odd number of wires, the remaining effective single SSH wire labeled by $l_0$ has the usual critical point at $\delta=0$, separating trivial and nontrivial phases. Therefore, the overall phase diagram consists of intersecting triangularly shaped gapless regions separated by insulating phases and, for an odd number of wires, an additional single critical line at $\delta=0$.

The insulating phases for ladders with an even number of legs are always trivial. For ladders with an odd number of legs, the insulating phases are trivial for $\delta>0$ and nontrivial, with $w=1$, for $\delta<0$. Figures~\ref{fig:TopologiaclPhases}(c) and (d) show the topological phase diagrams of six and seven perpendicularly coupled SSH wires, respectively. The transparent blue regions in Fig.~\ref{fig:TopologiaclPhases}(c) indicate gapless phases, while the transparent blue (pink) regions in Fig.~\ref{fig:TopologiaclPhases}(d) indicate gapless phases in which the effective single SSH wire is trivial (nontrivial).

\section{Topological characterization of critical and gapless phases}\label{sec:Critical_and_Gapless}
For a system with BDI symmetries, the conventional winding number $w$ is
well defined only when the bulk is gapped at zero energy. Using analytical
continuation of the off-diagonal block $h(k)$, we set $z=e^{ik}$ and introduce
$\mathcal{F}(z)=\det h(z)$. According to Verresen \textit{et al.}~\cite{ver18},
and provided $\mathcal{F}(z)$ is not identically zero, we
define the integer
\[
    \Omega=N_z-N_p
\]
which counts its zeros $N_z$ , with multiplicity, minus its poles $N_p$
strictly inside $|z|<1$. Zeros of
$\mathcal{F}(z)$ on the unit circle identify bulk zero-energy excitations. In
insulating phases $\Omega=w$, whereas in critical or gapless phases $w$ is ill
defined and $\Omega$ remains well defined. The exact decomposition established in
Sec.~\ref{sec:BandStructures} gives $\mathcal{F}(z)=\prod_s P_s(z)$ and hence $\Omega=\sum_s\Omega_s$, where $s$ labels an independent sector that preserves the
BDI symmetries.
Although $\Omega$ is additive, individual sector $s$ may retain
information about localized edge modes that is absent from the total.

\begin{figure}[t]
\centering
\includegraphics[width=0.4\textwidth]{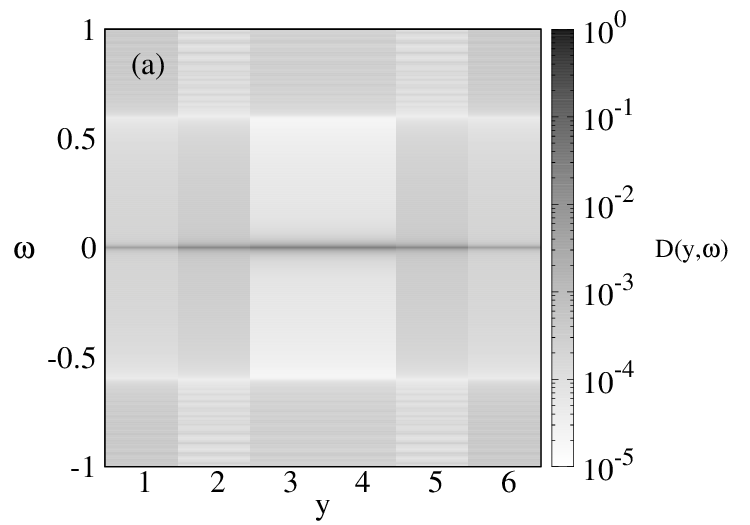}
\includegraphics[width=0.4\textwidth]{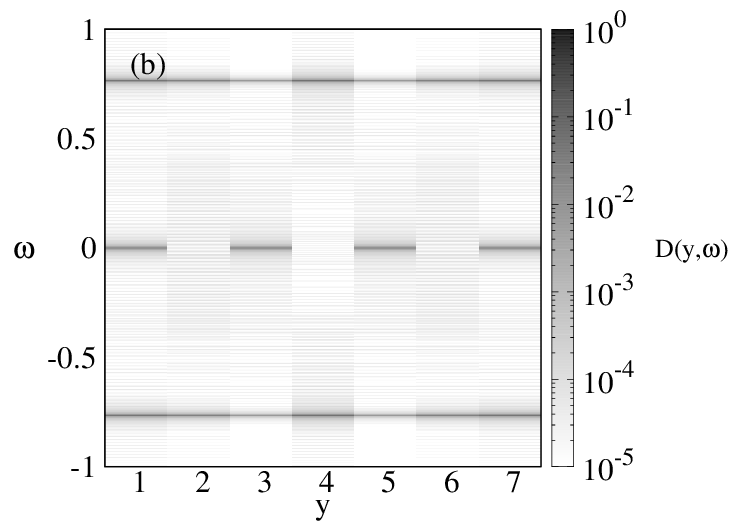}
\caption{Local density of states at the edges of (a) six diagonally coupled wires with $t_d=\tau^2_d$ and $\delta=0.3$ (b) seven perpendicularly coupled wires with $t_{\perp}=1$ and $\delta=-0.5$.
}
\label{fig:EdgeLDS}
\end{figure}

For the diagonally coupled SSH wires , we obtain
\begin{equation}\label{eq:PolyDiagSec5}
    P_{\mp l}(z)=t_{\mp l}+t'_{\mp l}z,
\end{equation}
for sectors $s=\mp l$ in addition to
\begin{equation}\label{eq:PolySingleSec5}
    P_{l_0}(z)=t+t'z
\end{equation}
for sector $s=l_0$ of odd $N_{\text{w}}$. There is one zero at
$z=-t_s/t'_s$ if $t'_s\ne0$, and no poles.
Consequently, a gapped effective SSH sector contributes $\Omega_s=1$
when $|t'_s|>|t_s|$ and $\Omega_s=0$ when $|t'_s|<|t_s|$. A sector with $|t'_s|=|t_s|\neq0$ has a zero on $|z|=1$ and contributes $\Omega_s=0$.
Therefore, the vertical critical lines $t_d=\tau_d^l$, at $\delta\ne0$
retain the contributions of the other gapped sectors. Depending on the winding numbers of the remaining gapped sectors, these critical lines may support exponentially localized zero-energy edge states, as illustrated in Fig.~\ref{fig:EdgeLDS}(a).

The decomposition into the effective SSH sectors reveals their shared band-
zone edge at $\pm\pi$. Thus, at $\delta=0$, away from the exceptional points discussed below, all effective SSH sectors become
simultaneously critical at $k=\pm\pi$, yielding $\Omega=0$ and no
exponentially localized zero-energy edge states.
The insulating phases on either side can nevertheless have different,
positive winding numbers. This shows that the winding numbers of the
neighboring insulating phases alone do not determine the existence of
localized edge modes at the intervening critical point. The result is consistent
with the edge-mode counting criterion of Ref.~\cite{ver18} for
$\Omega=0$, while qualifying a universal interpretation of its
concluding discussion of critical points between nontrivial phases.
Another interesting observation is at the exceptional intersections $\delta=0$ and $t_d^l=1$, where both $t_{-l}$ and $t^{\prime}_{-l}$
vanish identically, so $\mathcal{F}(z)\equiv0$ and $\Omega$ is undefined.
These points host zero-energy flat bands and must be excluded from the
zeros-minus-poles classification.

For the perpendicularly coupled wires, each effective $s=l$ subsystem gives
the Laurent polynomial
\begin{equation}\label{eq:LaurentPerpSec5}
    P_l(z)=\frac{tt'z^2+
    \bigl[t^2+t'^2-(t_\perp^l)^2\bigr]z+tt'}{z}.
\end{equation}
For $tt'>0$, each $P_l(z)$ has one pole at $z=0$.
Define $A_l=t^2+t'^2-(t_\perp^l)^2$ for $tt'>0$. The two zeros obey
$z_+z_-=1$, and their discriminant is
\begin{equation}\label{eq:Discriminant}
    \Delta_l=A_l^2-4t^2t'^2
\end{equation}
When $|A_l|>2tt'$, the zeros are real, with one strictly
inside and the other strictly outside the unit circle. Together with
the pole at $z=0$, they give $\Omega_l=0$. For
$-2tt'<A_l<2tt'$, the roots are distinct and lie on $|z|=1$. This corresponds
to the interior of the $l$-th gapless region,
$2|\delta|<t_\perp^l<2$. At $A_l=\pm2tt'$, the two roots merge on the
unit circle, leading to the boundaries $t_\perp^l=2|\delta|$ and
$t_\perp^l=2$ of the $l$-th gapless region. Since the two unit-circle zeros are excluded from $N_z$,
these critical and gapless cases have $\Omega_l=-1$. At $tt'=0$,
$P_l(z)=4-(t_\perp^l)^2$ is constant, with $\Omega_l=0$ unless
$t_\perp^l=2$. In the last exceptional case $P_l(z)\equiv0$ and the index is
undefined.

Let $m$ denote the number of critical or gapless effective perpendicular
two-wire subsystems, away from the exceptional points just identified.
For even $N_{\text{w}}$, the full index is $\Omega=-m$ in gapless and critical phases, and equals zero in all
gapped phases. For odd $N_{\text{w}}$, the remaining single-wire sector gives
$\Omega_{l_0}=1$ for $\delta<0$ and $\Omega_{l_0}=0$ for
$\delta\ge0$, so that
\[
  \Omega=\Omega_{l_0}-m.
\]
Thus for $\delta<0$ the total index can vanish or become negative even
though the $l_0$ sector remains topologically nontrivial. Indeed, the system
hosts localized edge mode for $\delta<0$. For example, in
Fig.~\ref{fig:EdgeLDS}(b), with
seven wires, $t_\perp=1$ and $\delta=-0.5$, two effective two-wire
subsystems are gapless ($m=2$), giving $\Omega=-1$, whereas
$\Omega_{l_0}=1$ and the localized edge mode persists.

This distinction concerns the protection of the independently conserved
transverse sectors. Ref.~\cite{ver18} already defines the
multiband invariant through a determinant and establishes its
additivity under stacking. Our result does not contradict those facts.
However, it demonstrates that the total invariant alone does not count all
localized modes when additional symmetries prevent sectors with
opposite index contributions from mixing. In particular, the $l_0$
mode is protected within the decoupled-sector structure, but its
persistence under arbitrary BDI-preserving perturbations that mix
transverse sectors is not implied by $\Omega$.

In the interiors of the perpendicular gapless regions, each gapless
two-wire subsystem provides two linearly dispersing Dirac channels.
Thus, with $m$ such subsystems, the central charge is $c=2m$; for odd
$N_{\text{w}}$, the critical $l_0$ sector adds one channel at $\delta=0$.
This counting concerns conformally invariant regions with linear
low-energy dispersion. At the boundaries of the triangular regions,
where roots merge and the dispersion is generically quadratic, and at
the exceptional zero-energy flat-band points, a conformal central
charge should not be assigned through the same counting procedure. This offers the
possibility of characterizing the conformally invariant critical and gapless
phases using the pair ($\Omega, c$) as introduced in~\cite{ver18}.

The exact decomposition of the coupled-wire Hamiltonians into independent effective sectors reflects their underlying transverse symmetries~\cite{del11,zha17,lin17,pad18}. For diagonally coupled wires, the common dimerization pattern leads to the simultaneous closing of all effective-sector gaps at $k=\pm\pi$ when $\delta=0$. For an odd number of perpendicularly coupled wires, the independent $l_0$ sector retains its nontrivial topology and localized edge modes for $\delta<0$, even when the full-system invariant vanishes or becomes negative. These observations highlight the importance of sector-resolved topological information and provide qualifications concerning the application of the edge-mode characterization of Ref.~\cite{ver18} to multiband systems with additional symmetries. They also motivate further investigation of how such symmetries modify the classification of critical and gapless phases.

\section{W-like edge states and confined coherent correlations:}
\label{sec:ZeroEdgeModes}
\begin{figure*}[t]
\centering
\includegraphics[width=0.4\textwidth]{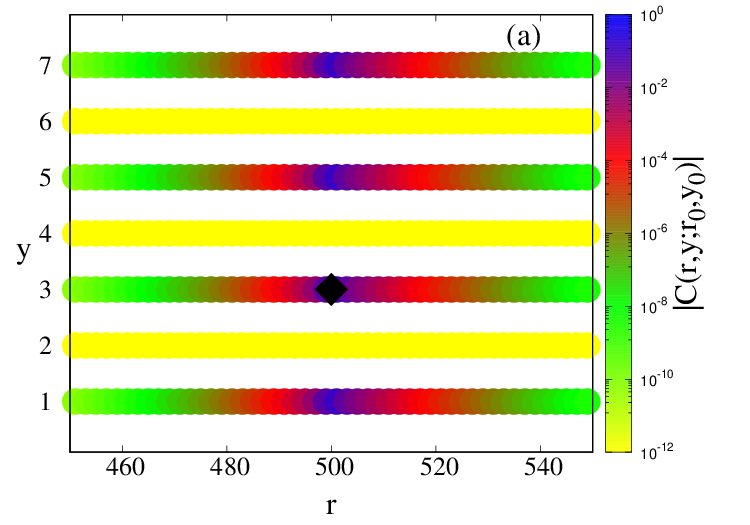}
\includegraphics[width=0.4\textwidth]{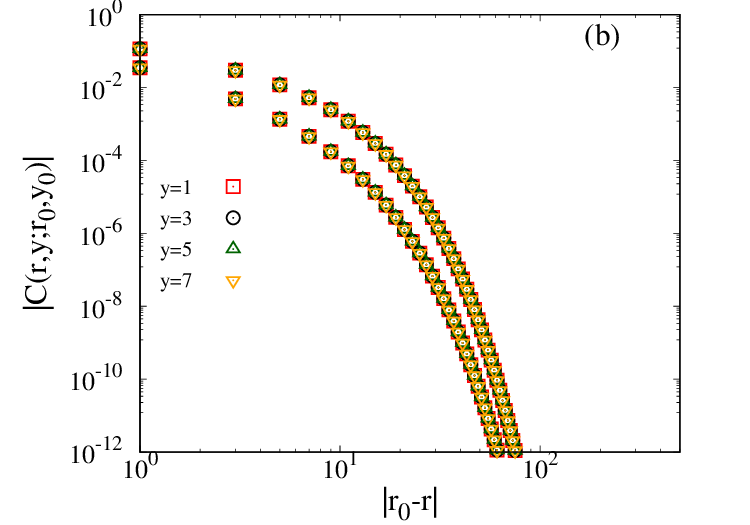}
\includegraphics[width=0.4\textwidth]{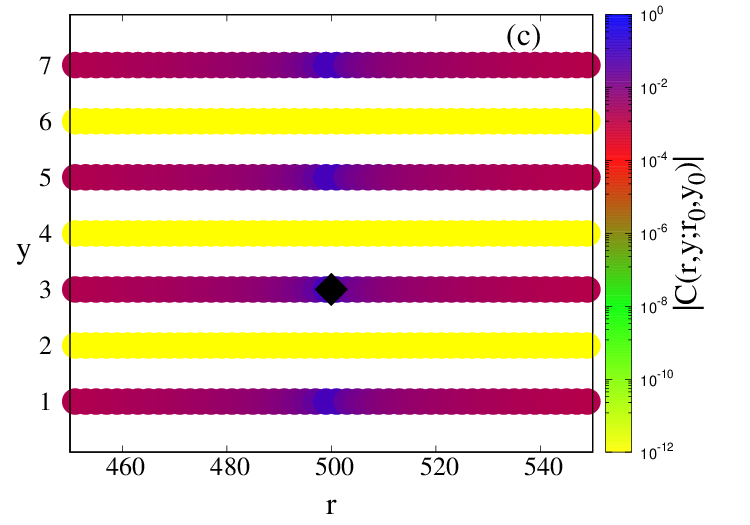}
\includegraphics[width=0.4\textwidth]{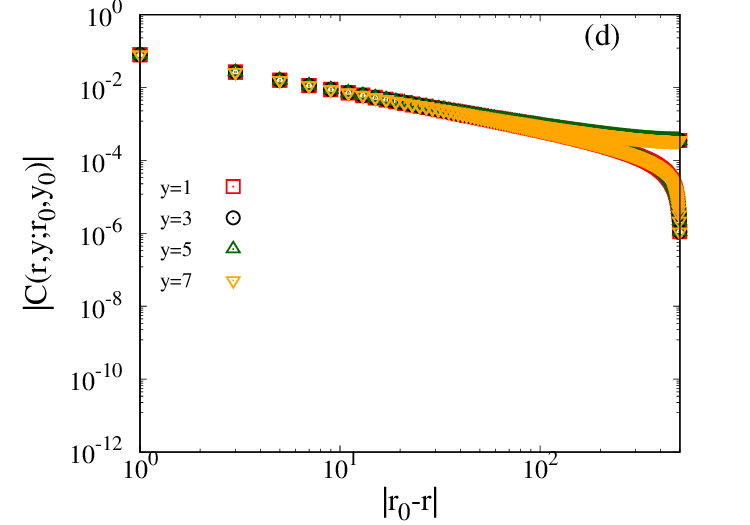}
\caption{Absolute values of the correlation function in seven perpendicularly coupled SSH wires with $L_x=1000$. The reference site, denoted by a black rhombus, is at $r_0=500$ and $y_0=3$. The perpendicular coupling is $t_{\perp}=6$, with (a),(b) $\delta=-0.3$ and (c),(d) $\delta=0$.}
\label{fig:CorrFunc}
\end{figure*}

The $k$-resolved single-particle correlation function is defined by
\begin{equation}\label{eq:k-resolved-correlation}
 C_{ij}(k)=\langle\Phi_{\mathrm{hf}}|c_{k,i}^{\dagger}c_{k,j}|\Phi_{\mathrm{hf}}\rangle
=\sum_{\mu\in\mathrm{occ}}\phi_\mu^*(k;i)\phi_\mu(k;j),
\end{equation}
Here, $i=(x,y)$, $|\Phi_{\mathrm{hf}}\rangle=\prod_k\prod_{\mu\in\mathrm{occ}}d_{k,\mu}^{\dagger}|\Phi\rangle$ is the ground state at half filling given as Slater determinant, with $d_{k,\mu}^{\dagger}=\sum_i\phi_\mu(k;i)c_{k,i}^{\dagger}$ and $|\Phi\rangle$ is the vacuum state. As stated above, each $\mu=\mp l$ label corresponds to bands of an effective single SSH wire. Therefore, the corresponding eigenvectors form a complete basis for the effective SSH-wire subsystem. For an odd number of perpendicularly coupled wires at half filling, all states in the $-l$ shifted sectors are completely filled, while all states in the $+l$ shifted sectors are empty for $t_{\perp}>\tau^N_{\perp}$. For each $l$, the two bands of a completely filled shifted sector form a complete basis within that sector. Then one has $\sum_{s\in\{-l\}}|\phi_{s}(k)\rangle \langle\phi_{s}(k)|=\mathbb{1}_{-l}$, where $\{-l\}$ are the two bands inside a fixed $-l$ sector and $\mathbb{1}_{-l}$ is the identity restricted to the $-l$ effective SSH subsystem. These completely filled sectors therefore contribute only local terms after Fourier transformation and do not contribute to $C_{x,x_0;y,y_0}(u-u_0)$ for $u-u_0\neq0$. Consequently, the long-distance single-particle correlation function comes exclusively from the projector of the occupied lower band of the unshifted $l_0$ sector.
Then, as discussed in Sec.~\ref{sec:BandStructures}, the structure of the eigenvector
\begin{equation}\label{eq:phi_l0}
|\phi_{l_0}\rangle=
 \begin{pmatrix}
1 & 0 & -1 & 0 & \hdots & -\frac{T(k)}{|T(k)|} & 0 & \frac{T(k)}{|T(k)|} & 0 & \hdots
\end{pmatrix}^{\mathrm T}
\end{equation}
has important consequences for the correlation function. The real-space-resolved correlation is given, for large $N_u$, by the transformation
\begin{equation}\label{eq:real-space-resolved-correlation}
 C_{x,x_0;y,y_0}(u-u_0) = \frac{1}{2\pi}\int^{\pi}_{-\pi} e^{-ik(u-u_0)} C_{x,x_0;y,y_0}(k) dk.
\end{equation}
Therefore, if $y$ or $y_0$ is even, then $C_{x,x_0;y,y_0}(u-u_0)\propto\delta_{u,u_0}$. Otherwise, for any $x_0$ and odd $y_0$, the correlation decays coherently in all legs with odd index $y$, exactly as in the single SSH chain, up to a normalization factor, i.e., with exponential decay for $\delta\neq0$ and with power-law decay for $\delta=0$. This behavior persists for coupled wires of finite length. Figures~\ref{fig:CorrFunc}(a) and~\ref{fig:CorrFunc}(c) show, in color-density plots, how the correlation propagates coherently in all wires with odd $y$ indices. The system consists of seven perpendicularly coupled SSH wires with $L_x=1000$ and $t_{\perp}=6$. The reference site is $(r_0=\frac{L_x}{2},y_0=3)$. In Fig.~\ref{fig:CorrFunc}(b), the correlations decay exponentially for $\delta=-0.3$. More interestingly, the coherent decay is preserved at $\delta=0$, but with algebraic decay, as shown in Fig.~\ref{fig:CorrFunc}(d).

The alternating pattern found in $|\phi_{l_0}\rangle$ is also reflected in the apparently equal distribution of the LDOS among the edges with odd $y$ indices, as seen, for instance, in Fig.~\ref{fig:EdgeLDS}(b). In the nontrivial gapped phases with $t_{\perp}>\tau^N_{\perp}$, the LDOS vanishes in the wires with even $y$ indices, but remains exactly equally distributed among the wires with odd $y$ indices. Thus, the single-particle edge states $|\phi_{\alpha^{\prime}}\rangle$ in such perpendicularly coupled SSH wires take the form
\begin{equation}\label{eq:l0-x}
|\phi_{\alpha^{\prime}}\rangle = \frac{1}{\sqrt{N+1}} \sum_{r} \alpha_{r} \sum^{N+1}_{m=1} \left(-1\right)^{m+1} c^{\dag}_{r,2m-1} |\Phi\rangle ,
\end{equation}
For instance, the projection of $|\phi_{\alpha^{\prime}}\rangle$ onto rung $r$ for seven perpendicularly coupled SSH wires is given by
\begin{equation}\label{eq:l0-x-SevenWires}
|\alpha^{\prime}_{r}\rangle = \frac{1}{\sqrt{N+1}} \alpha_{r} \left[|1,0,0,0,0,0,0\rangle - |0,0,1,0,0,0,0\rangle \right. 
+ \left. |0,0,0,0,1,0,0\rangle - |0,0,0,0,0,0,1 \rangle \right].
\end{equation}
The equally distributed probabilities $|\alpha_{r}|^2/\left(N+1\right)$ decay exponentially along the $x$ direction for $\delta<0$. If rung $r$ is adiabatically disconnected from the rest of the system, it forms a state resembling a W state~\cite{dur00}.

Hence, an entangled edge mode emerges from a noninteracting system, in the sense of mode entanglement distributed across the odd-indexed wires, forming a W-like state in the single-particle sector.
However, we have to emphasize that this W-like state is a coherent single-particle superposition across spatial modes but not a genuine multipartite entanglement between particles.
Such coherent behavior should have important consequences for the coherence of real-time dynamics and for the transport properties of perpendicularly coupled SSH wires. One may therefore expect the $l_0$ sector to contribute to coherent transport along the odd-indexed wires, while the other sectors may also contribute to transport. The situation becomes more intricate in the gapless and insulating phases at $t_{\perp}<\tau^N_{\perp}$. However, for $t_{\perp}>\tau^N_{\perp}$, the decay of the correlation function is not affected by reducing $t_{\perp}$ until the critical line $\tau^N_{\perp}$ is crossed.

\section{Conclusion and outlook:}\label{sec:conclusion}
We identified the global half-filled phase diagrams for an arbitrary number of diagonally or perpendicularly coupled SSH wires. To the best of our knowledge, this constitutes the first analytically exact determination of these phase diagrams for both coupling geometries at arbitrary finite widths. Despite the simplicity of the underlying model, these systems provide a versatile platform for exploring not only topological insulating phases, as in the single SSH wire, but also critical and gapless phases. We further characterized these critical and gapless phases using the generalized topological index, revealing the importance of independent effective sectors for understanding their topological properties and localized edge modes. Finally, we identified W-like single-particle edge states and confined coherent correlations in odd perpendicularly coupled wires.

Our analytical treatment exploits the exact solutions of modified Toeplitz-plus-Hankel matrices, to which the relevant Hamiltonian matrices belong. This class of matrices encompasses
a wide range of structures beyond the specific forms considered here,
including different coupling patterns. As a result, the present approach provides a systematic framework for
determining band structures and phase diagrams for an arbitrary number
of coupled wires, and suggests that similar techniques can be applied to
a broader class of systems whose Hamiltonians, or functions thereof,
share a modified Toeplitz-plus-Hankel structure.
The resulting decomposition into independent effective subsystems
provides a natural framework, reminiscent of approaches used in 
weakly interacting ladder systems~\cite{led00}, and provides a noninteracting reference for investigating the corresponding interacting coupled-wire systems.

Another key result is the emergence of confined coherent correlations in perpendicularly coupled SSH wires with an odd number of legs for sufficiently strong perpendicular coupling. These nonlocal correlations extend coherently along wires with odd indices while vanishing on even-indexed wires. This behavior and the formation of mode-entangled edge states resembling W-like states in the single-particle sector both originate from the structure of the underlying eigenmodes. The consequences of these coherent structures for transport properties and nonequilibrium dynamics, as well as their fate under interactions, remain to be investigated~\cite{bar18,gho23}. The W-like structure of the edge modes may also be of interest for applications requiring coherent multichannel outputs, such as quantum random number generation.

The experimental realization of these coupled SSH wires is possible in a variety of setups, similar to those used for experimentally realizing single SSH wires and a few coupled SSH wires. One possible route is the engineering of atomic lattices on surfaces, using scanning tunneling microscopy to create a lattice of vacancies in a chlorine monolayer on top of a Cu(100) surface~\cite{dro17,kha19,hud20}. In such setups, one may expect the detection of edge modes and domain walls that would confirm the findings of this work. Additionally, the ability to introduce a potential bias between the edges of engineered coupled wires may provide useful insight into the coherent correlations in systems with an odd number of perpendicularly coupled wires. The presence of flat bands in diagonally coupled wires can make the system susceptible to interaction effects~\cite{bis11,cao18}, potentially leading to exotic quantum states in experimental realizations.

For a single SSH wire, there is an established adiabatic correspondence between its topological phases and the symmetry-protected topological (SPT) phases of the one-dimensional SSH-Hubbard model at half filling and the dimerized spin-$\frac{1}{2}$ Heisenberg chain~\cite{hid92,man12,mik22}. Knowledge of the phase diagrams for an arbitrary number of coupled SSH chains can facilitate the understanding of possible analogous correspondences in coupled SSH-Hubbard and dimerized Heisenberg ladders with corresponding lattice structures, dimerization patterns, and inter-wire couplings. The topological phases of corresponding dimerized spinless fermion $t$-$V$ ladders are also of interest, considering previous investigations of interacting SSH ladders with different dimerization patterns~\cite{nersesyan20,parida26}. The well-known behavior of uniform systems with vanishing dimerization~\cite{lin97,dag96,giamarchi04}, together with previous investigations of dimerized Heisenberg ladders~\cite{totsuka95,gibson11} and topologically distinct phases in spin ladders~\cite{kim00}, should further support such an effort. In particular, the established odd-even behavior of perpendicularly coupled Hubbard and Heisenberg ladders motivates investigating possible analogous correspondences in their dimerized counterparts, while the corresponding diagonally coupled systems remain less systematically understood. However, the complexity of coupled wires, together with the possible competition between different extreme limits, can lead to richer behavior~\cite{abd25}. Moreover, it is interesting to understand the manifestation of spin-charge separation and the evolution of the spin and charge excitation gaps in the corresponding SSH-Hubbard ladders~\cite{montorsi17}. Investigations in this direction are ongoing.

Finally, it is well established that ground states of systems in certain strictly one-dimensional nontrivial SPT phases can serve as resources for measurement-based quantum computation~\cite{bre08,miy10,els12,mil15,ste17,rau23,adh24}. The use of quasi-one-dimensional systems for this purpose is also an active research direction~\cite{wei12,ste19}. Therefore, identifying SPT phases in coupled SSH-Hubbard and dimerized Heisenberg ladders may be useful for realizing such resources.

\section*{Acknowledgements}
We thank Q. Deng for the useful correspondence regarding solutions of Toeplitz-plus-Hankel matrices and nonsymmetric matrices. We thank E. Jeckelmann, R. Rausch and L. Santos for the useful comments on the manuscript. We thank G. Platero, S. Rachel and P. Herringer for the useful discussions.
\bibliography{bibliography.bib}


\section*{Disclosure of generative AI use}
OpenAI's ChatGPT was used to assist with language editing, manuscript revisions, literature searches, and mathematical consistency checks and corrections. All AI-assisted suggestions were critically reviewed by the author, who takes full responsibility for the scientific content and final manuscript.
\end{document}